\documentclass[prd,preprint,superscriptaddress,amsmath,amssymb,nofootinbib]{revtex4}
\usepackage{graphicx}
\usepackage{dcolumn}
\usepackage{bm}
\usepackage{amssymb}
\usepackage{amsmath}
\usepackage{epsfig}    
\usepackage{color}
\usepackage{slashed}
\usepackage{hhline}
\usepackage{tikz-feynman,contour}
\def\be{\begin{equation}}
\def\ee{\end{equation}}
\newcommand{\bea}{\begin{eqnarray}}
\newcommand{\eea}{\end{eqnarray}}
\newcommand{\nn}{\nonumber}

\numberwithin{equation}{section}

\begin{document}

 \begin{flushright} {CTP-SCU/2021021}, APCTP Pre2021-013 \end{flushright}

\title{Quark and lepton flavor model with leptoquarks in a modular $A_4$ symmetry}

\author{Takaaki Nomura}
\email{nomura@scu.edu.cn}
\affiliation{College of Physics, Sichuan University, Chengdu 610065, China}

\author{Hiroshi Okada}
\email{hiroshi.okada@apctp.org}
\affiliation{Asia Pacific Center for Theoretical Physics, Pohang 37673, Republic of Korea}
\affiliation{Department of Physics, Pohang University of Science and Technology, Pohang 37673, Republic of Korea}

\author{Yuta Orikasa}
\email{Yuta.Orikasa@utef.cvut.cz}
\affiliation{Institute of Experimental and Applied Physics, Czech Technical University in Prague, 110 00 Prague 1, Czech Republic}

\date{\today}

\begin{abstract}
We propose a quark-lepton model via leptoquarks and modular $A_4$ symmetry.
Since the neutrino mass is induced at one-loop level mediated by down quarks as well as leptoquarks, we have to explain lepton and quark masses and mixings with a single modulus $\tau$.
Here, we find predictions for lepton and quark sectors with unified modulus $\tau$, and show several constraints originating from leptoquarks.
\end{abstract}

\maketitle

\section{Introduction}
Since lepto-quark(LQ) bosons connect lepton and quark sectors, these models potentially explain several new physics beyond the standard model (SM); {\it e.g.}, lepton(muon or electron) anomalous magnetic dipole moment ($g-2$)~\cite{Bauer:2015knc, Chen:2016dip, Chen:2017hir, Nomura:2021oeu,ColuccioLeskow:2016dox}, $B$ meson decays such as $b\to s\mu\bar\mu$~\cite{Sahoo:2015wya,Becirevic:2016yqi,Chen:2016dip, Chen:2017hir,Nomura:2021oeu,Cai:2017wry} and $b\to c\ell\bar\nu_\ell$($\ell=e,\mu,\tau$)~\cite{Sakaki:2013bfa,Becirevic:2016yqi,Chen:2017hir,Cai:2017wry}~\footnote{The anomaly of $b\to c\ell\bar\nu_\ell$ processes are observed in experiments~\cite{Huschle:2015rga, Lees:2012xj,Lees:2013uzd,Hirose:2016wfn,Abdesselam:2016cgx,Aaij:2015yra,Aaij:2017deq}, and LQ model is one of the most promising explanations on this anomaly.}, and nonzero neutrino masses~\cite{AristizabalSierra:2007nf, Cheung:2016fjo,Cai:2017wry}.
Especially, muon $g-2$ anomaly is recently reported by E989 experiment at Fermilab combining BNL result~\cite{Abi:2021gix}, and its value is deviated from the SM by 4.2$\sigma$ as follows:
\begin{align}
\Delta a_\mu = (25.1\pm 5.9)\times 10^{-10}. 
\end{align}
Also, the LHCb collaboration~\cite{Aaij:2021vac} recently reported anomaly of rare $B$ meson decays of $b\to s\mu\bar\mu$ that is understood as violation of lepton universality. The updated result is given by
\begin{align}
\frac{BR(B^+\to K^+\mu^-\mu^+)}{BR(B^+\to K^+e^-e^+)}= 0.846_{-0.039-0.012}^{+0.042+0.013}\quad (1.1 {\rm GeV}^2 < q^2 < 6 {\rm GeV}^2), 
\end{align}
where first(second) uncertainty is statistical(systematic) one and $q^2$ is the invariant mass squared for dilepton.
In addition to the above phenomenologies, interestingly, the nonzero Majorana neutrino mass at one-loop can be realized without any additional symmetries by introducing appropriate LQs~\cite{Cheung:2016fjo}. This may be natural realization of tiny neutrino mass model
due to loop suppression.

Considering above issues, one finds that Yukawa flavor structure is also very important to explain them.
Recently, powerful symmetries to restrict the number of parameters in Yukawa couplings, so called "modular flavor symmetries", were proposed by authors in refs.~\cite{Feruglio:2017spp,deAdelhartToorop:2011re}, in which 
they have applied modular originated non-Abelian discrete flavor symmetries to quark and lepton sectors.
One remarkable advantage of applying this symmetries is that dimensionless couplings of model can be transformed into non-trivial representations under those symmetries, and all the dimensionless values are uniquely fixed once modulus is determined in fundamental region. We then do not need the scalar fields to obtain a predictive mass matrix.
Along the line of this idea, a vast reference has recently appeared in the literature, {\it e.g.},  $A_4$~\cite{Feruglio:2017spp, Criado:2018thu, Kobayashi:2018scp, Okada:2018yrn, Nomura:2019jxj, Okada:2019uoy, deAnda:2018ecu, Novichkov:2018yse, Nomura:2019yft, Okada:2019mjf,Ding:2019zxk, Nomura:2019lnr,Kobayashi:2019xvz,Asaka:2019vev,Zhang:2019ngf, Gui-JunDing:2019wap,Kobayashi:2019gtp,Nomura:2019xsb, Wang:2019xbo,Okada:2020dmb,Okada:2020rjb, Behera:2020lpd, Behera:2020sfe, Nomura:2020opk, Nomura:2020cog, Asaka:2020tmo, Okada:2020ukr, Nagao:2020snm, Okada:2020brs, Yao:2020qyy, Chen:2021zty, Kobayashi:2021bgy, Kashav:2021zir, Okada:2021qdf},
$S_3$ \cite{Kobayashi:2018vbk, Kobayashi:2018wkl, Kobayashi:2019rzp, Okada:2019xqk, Mishra:2020gxg, Du:2020ylx},
$S_4$ \cite{Penedo:2018nmg, Novichkov:2018ovf, Kobayashi:2019mna, King:2019vhv, Okada:2019lzv, Criado:2019tzk,
Wang:2019ovr, Zhao:2021jxg, King:2021fhl, Ding:2021zbg, Zhang:2021olk, gui-jun},
$A_5$~\cite{Novichkov:2018nkm, Ding:2019xna,Criado:2019tzk}, double covering of $A_5$~\cite{Wang:2020lxk, Yao:2020zml, Wang:2021mkw}, larger groups~\cite{Baur:2019kwi}, multiple modular symmetries~\cite{deMedeirosVarzielas:2019cyj}, and double covering of $A_4$~\cite{Liu:2019khw, Chen:2020udk}, $S_4$~\cite{Novichkov:2020eep, Liu:2020akv}, and the other types of groups \cite{Kikuchi:2020nxn, Almumin:2021fbk, Ding:2021iqp, Feruglio:2021dte, Kikuchi:2021ogn, Novichkov:2021evw} in which masses, mixing, and CP phases for the quark and/or lepton have been predicted~\footnote{For interest readers, we provide some literature reviews, which are useful to understand the non-Abelian group and its applications to flavor structure~\cite{Altarelli:2010gt, Ishimori:2010au, Ishimori:2012zz, Hernandez:2012ra, King:2013eh, King:2014nza, King:2017guk, Petcov:2017ggy}.}.
Moreover, a systematic approach to understand the origin of CP transformations has been discussed in Ref.~\cite{Baur:2019iai}, 
and CP/flavor violation in models with modular symmetry was discussed in Refs.~\cite{Kobayashi:2019uyt,Novichkov:2019sqv,Kobayashi:2021bgy,1869542}, 
and a possible correction from K\"ahler potential was discussed in Ref.~\cite{Chen:2019ewa}. Furthermore,
systematic analysis of the fixed points (stabilizers) has been discussed in Ref.~\cite{deMedeirosVarzielas:2020kji}.
A very recent paper of Ref.~\cite{Ishiguro:2020tmo} finds a favorable fixed point $\tau=\omega$ among three fixed points, which are the fundamental domain of PSL$(2,Z)$, by systematically analyzing the stabilized moduli values in the possible configurations of flux compactifications as well as investigating the probabilities of moduli values.
{It is then interesting to discuss a LQ model under the framework of modular flavor symmetry 
since we are motivated to consider lepton and quark sector together as a LQ connect these sectors and some predictions in both sector can be expected.
}

In this paper, we focus on the quark and lepton masses and  mixings based on a LQ model in ref.~\cite{Cheung:2016fjo}, introducing  modular $A_4$ symmetry to reduce free parameters of Yukawa couplings. 
Since the quark sector connects to the lepton sector via LQ, charge assignments for quarks(leptons) directly affect the leptons(quarks).
In this sense, it would be a good motivation towards unification of quark and lepton flavor in $A_4$ modular symmetry.

This paper is organized as follows.
In Sec.~II, we review our model of quark and lepton.
In Sec.~III, we have numerical analysis and show several  results for normal and inverted hierarchies.
We conclude in Sec.~IV. In appendix, we summarize several features of modular $A_4$ symmetry.

\section{Model}

\begin{table}[t!]
\begin{tabular}{|c||c|c|}
\hline\hline  
                   & ~$\eta$~  & ~$\Delta$~ \\\hline 
$SU(3)_C$ & $\bm{3}$  & $\bar{\bm3}$  \\\hline 
$SU(2)_L$ & $\bm{2}$  & $\bm{3}$  \\\hline 
$U(1)_Y$   & $\frac16$ & $\frac13$    \\\hline
$A_4$   & $\bm{1}$ & $\bm{1}$    \\\hline
$-k_I$   & $-2$ & $-2$    \\\hline
\end{tabular}
\caption{\small 
Charge assignments of the LQ bosons $\eta$ and $\Delta$  
 under $SU(3)_C \times SU(2)_L\times U(1)_Y\times A_4$ where $k_I$ is the number of modular weight.}
\label{tab:1}
\end{table}

\begin{center} 
\begin{table}[t!]
\begin{tabular}{|c||c|c|c|c|c|c|}\hline\hline  
  & \multicolumn{5}{c|}{Fermions}   \\ \hline \hline
& ~$Q_L$~& ~$\bar u_R$~  & ~$\bar d_R$~& ~$L_L$~& ~$\bar e_R$~    \\\hline\hline 
$SU(3)_C$ & $\bm{3}$  & $\bar{\bm{3}}$ & $\bar{\bm{3}}$ & $\bm{1}$ & $\bm{1}$ \\\hline 
$SU(2)_L$ & $\bm{2}$  & $\bm{1}$  & $\bm{1}$  & $\bm{2}$  & $\bm{1}$ \\\hline 
$U(1)_Y$   & $\frac16$ & $-\frac23$ & $\frac13$ & $-\frac12$  & $1$     \\\hline
$A_4$ & $\bm{3}$ & $\bm{1,1'',1'}$ & $\bm{1,1'',1'}$ & $\bm3$ & $\bm{1,1'',1'}$     \\ \hline
$-k_I$ & $-2$ & $-4$ & $0$ & $-2$ & $0$   \\
\hline
\end{tabular}
\caption{Charge assignments of the SM fermions under $SU(3)_C \times SU(2)_L\times U(1)_Y\times A_4$ where $k_I$ is the number of modular weight.}
\label{tab:fields-inverse}
\end{table}
\end{center}

In this section, we review our model.
It is known that introducing proper leptoquarks lead us to a radiative seesaw model without any additional symmetries such as $Z_2$. Here, we introduce two types of leptoquarks $\eta$ and $\Delta$ based on Ref.~\cite{Cheung:2016fjo}.
The color-triplet $\eta$ has $SU(2)_L$ doublet with $1/6$ hypercharge, 
and the color-antitriplet $\Delta$ has $SU(2)_L$ triplet with 
$1/3$ hypercharge, where these new bosons and their charges are summarized in Table~\ref{tab:1}.
Then, the valid Lagrangian to induce the quark and lepton mass matrices is given by
\begin{align}
-\mathcal{L}_{Y}^{q}
&= y^u_{ij} \bar u_{R_i}  (i\sigma_2) H^* Q_{L_j} +  y^d_{ij} \bar d_{R_i} H Q_{L_j} + {\rm h.c.},
\label{Eq:lag-quark}\\
-\mathcal{L}_{Y}^{\ell}
&= h_{ij} \bar e_{R_i} H L_{L_j} + {\rm h.c.}.
\label{Eq:lag-lepton}
\end{align}
The Lagrangian for the mixing between the quark and lepton and nontrivial potential are given by  
\begin{align}
-\mathcal{L}_{Y}^{mix}
&= f_{ij} \overline{d_{R_i}}  \eta^T  (i\sigma_2) L_{L_j} + g_{ij} \overline{ Q^c_{L_i}}
 (i\sigma_2) \Delta L_{L_j} + {\rm h.c.},\\
\mathcal{V}
&\supset -\mu H^\dag \Delta \eta+ {\rm h.c.},
\label{Eq:lag-flavor}
\end{align}
where $(i,j)=1-3$ are family indices, $\sigma_2$ is the second 
Pauli matrix, and $H$ is the SM Higgs field that develops a 
nonzero VEV, which is symbolized by 
$\langle H\rangle\equiv v/\sqrt2\approx 246/\sqrt2$ GeV, and $H$ has nonzero modular weight.
Here, we parameterize components of the scalars as follows: 
\begin{align}
&H =\left[
\begin{array}{c}
w^+\\
\frac{v+\phi+iz}{\sqrt2}
\end{array}\right],\quad 
\eta =\left[
\begin{array}{c}
\eta_{2/3}\\
\eta_{-1/3}
\end{array}\right],\quad 
\Delta =\left[
\begin{array}{cc}
\frac{\delta_{1/3}}{\sqrt2} & \delta_{4/3} \\
\delta_{-2/3} & -\frac{\delta_{1/3}}{\sqrt2}
\end{array}\right],
\label{component}
\end{align}
where the subscript of the fields represents the electric charge, and $w^+$ and $z$ are absorbed by the 
longitudinal component of the $W^+$ and $Z$ bosons, respectively.
Due to the $\mu$ term in Eq.~(\ref{Eq:lag-flavor}), the charged components
with $1/3$ and $2/3$ electric charges mix each other.
Here, we parametrize their mixing matrices and mass eigenstates as follows: 
\begin{align}
&\left[\begin{array}{c} \eta_{i/3} \\ \delta_{i/3} \end{array}\right] = 
O_i \left[\begin{array}{c} A_i \\ B_i \end{array}\right],\quad
O_i\equiv 
\left[\begin{array}{cc} c_{a_i} & s_{a_i} \\
 -s_{a_i} & c_{a_i}   \end{array}\right], \quad (i=1,2),
\end{align}
where their masses are denoted as $m_{A_i}$ and  $m_{B_i}$ respectively. 
The interactions in terms of the mass eigenstates can be written as 
\begin{align}
- \mathcal{L}_{Y}^q \approx & \ m_{u_{ij}} \bar u_{R_i}  u_{L_j} +  m_{d_{ij}} \bar d_{R_i}  d_{L_j}+ {\rm h.c.},
\label{eq:quark}\\
 - \mathcal{L}_{Y}^\ell \approx & \ m_{\ell_{ij}}\bar e_{R_i}  \ell_{L_j} + {\rm h.c.},\\
 - \mathcal{L}_{Y}^{mix} \approx & \ 
	f_{ij} \overline{ d_{R_i}} \nu_{L_j} (c_{a_1} A_1^* +s_{a_1} B_1^*)
	-\frac{g_{ij}}{\sqrt2} \overline{ d_{L_i}^c} \nu_{L_j} (-s_{a_1} A_1 + c_{a_1} B_1) 
\label{eq:neut}
\\
& -f_{ij} \overline{d_{R_i}} \ell_{L_j} (c_{a_2} A_2 +s_{a_2} B_2)
	-\frac{g_{ij}}{\sqrt2} \overline{u_{L_i}^c} \ell_{L_j} (-s_{a_1} A_1 + c_{a_1} B_1) 
\label{eq:lfvs-1}\\
&
- {g_{ij}} \overline{d_{L_i}^c} \ell_{L_j}\delta_{4/3} \;
 + {g_{ij}} \overline{u_{L_i}^c} \nu_{L_j} (-s_{a_2} A_{2}^* + c_{a_2} B_{2}^*) + {\rm h.c.},
\label{eq:lfvs-2}
\end{align}
where we define $m_{u_{ij}}\equiv \frac{v y^u_{ij}}{\sqrt2} $, $m_{d_{ij}}\equiv \frac{v y^d_{ij}}{\sqrt2} $, and $m_{\ell_{ij}}\equiv \frac{v h_{ij}}{\sqrt2} $.

The next task is to determine the matrices of $y^u,y^d,h,f,g$ via modular $A_4$ symmetry.
In the quark sector, we assign $Q_L$ to be ${\bf 3}$ and $-2$, $\bar u_R$ to be $\{\bf 1,1'',1'\}$ and $-4$, and $\bar d_R$ to be $\{{\bf 1,1'',1'}\}$ and $0$ under $A_4$ and $-k$, respectively.
This assignment is the same as the one in ref.~\cite{Okada:2020rjb}, and it is already known that allowed region~\cite{Okada:2019uoy}. 
 Thus, we will work on the same  $\tau$ region of the lepton sector in our numerical analysis. 
  The up-type quark mass matrix is written as:
\begin{align}
\begin{aligned}
y^u=
\begin{pmatrix}
a_u & 0 & 0 \\
0 &a_c & 0\\
0 & 0 &a_t
\end{pmatrix} \left [
\begin{pmatrix}
f_1  & f_3 & f_2  \\
f_2  & f_1  & f_3 \\
f_3  & f_2 &  f_1 
\end{pmatrix}
+ 
\begin{pmatrix}
g_{u1} & 0 & 0 \\
0 &g_{u2} & 0\\
0 & 0 &g_{u3}
\end{pmatrix}
\begin{pmatrix}
f'_1  & f'_3 & f'_2  \\
f'_2  & f'_1  & f'_3  \\
f'_3  & f'_2 &  f'_1 
\end{pmatrix}
\right ],
\end{aligned}
\label{matrix6}
\end{align}
where  
$Y^{(6)}_{3}\equiv[f_1,f_2,f_3]^T$ and $Y^{(6)}_{3'}\equiv[f'_1,f'_2,f'_3]^T$, $g_{u1}=\alpha'_u/\alpha_u$, $g_{u2}=\beta'_u/\beta_u$
and $g_{u3}=\gamma'_u/\gamma_u$ are complex parameters,
and $a_u$,  $a_c$ and  $a_t$ can be used to fit the masses of up quarks.
The explicit forms of $f_i$ and $f'_i$ are summarized in Appendix.  
Then $m_u$ is diagonalized by two unitary matrices as $D_u=V_{u_R}^\dag m_u V_{u_L}$, where $D_u\equiv {\rm diag}(m_u,m_c,m_t)$ is mass eigenvalues.
Therefore, we find $|D_u|^2=V^\dag_{u_L} m_u^\dag m_u V_{u_L}$.
 On the other hand, the down-type quark mass matrix is given as:  
 \begin{align}
 &\begin{aligned}
 y^d= 
 \begin{pmatrix}
 a_d & 0 & 0 \\
 0 &a_s & 0\\
 0 & 0 &a_b
 \end{pmatrix}
 \begin{pmatrix}
 y_1 & y_3& y_2\\
 y_2 & y_1 & y_3 \\
 y_3 & y_2&  y_1
 \end{pmatrix} \,,
 \end{aligned} 
 \label{down}
 \end{align}
 where and $a_d$,  $a_s$ and  $a_b$ can be used to fit the masses of down quarks,
 and $Y^{(2)}_{3}\equiv[y_1,y_2,y_3]^T$ in Appendix.
Then $m_d$ is diagonalized by two unitary matrices as $D_d=V_{d_R}^\dag m_d V_{d_L}$, where $D_d\equiv {\rm diag}(m_d,m_s,m_b)$ is mass eigenvalues.
Therefore, we find $|D_d|^2=V^\dag_{d_L} m_d^\dag m_d V_{d_L}$.
Finally, we get the observable mixing matrix $V_{CKM}$ as follows:
\begin{align}
V_{CKM}= V^\dag_{u_L} V_{d_L}.
\end{align}

\subsection{Lepton sector}
Now let us move on the lepton sector.
We assign $L_L$ to be ${\bf 3}$ and $-2$ and $\bar e_R$ to be $\{\bf 1,1'',1'\}$ and $0$ under $A_4$ and $-k$, respectively.
Here, both of the leptoquark scalars are assigned to be true $A_4$ singlets with $-2$ modular weight. 
The assignments of $A_4$ and $-k$ are also summarized in Tables~\ref{tab:1} and \ref{tab:fields-inverse}.
Under these assignments, we can write down the concrete matrices as follows:

\begin{align}
h&=
\left[\begin{array}{ccc}
a_\ell &0 &0 \\
0 &b_\ell &0 \\
0 &0 &c_\ell \\
\end{array}\right] 
\left[\begin{array}{ccc}
y_1 & y_3 &y_2 \\
y_2 &y_1 &y_3 \\
y_3 &y_2 &y_1 \\
\end{array}\right] ,\\
f&=
\left[\begin{array}{ccc}
a_\eta &0 &0 \\
0 &b_\eta &0 \\
0 &0 &c_\eta \\
\end{array}\right] 
\left[\begin{array}{ccc}
y'_1 & y'_3 &y'_2 \\
y'_2 &y'_1 &y'_3 \\
y'_3 &y'_2 &y'_1 \\
\end{array}\right] ,\\
g&=
a Y^{(6)}_1\left[\begin{array}{ccc}
1 &0 &0 \\
0 &0 &1 \\
0 &1 &0 \\
\end{array}\right] 
+\frac{b}{3}
\left[\begin{array}{ccc}
2 f_1 &- f_3& - f_2 \\
- f_3 & 2 f_2 & - f_1 \\
- f_2 &- f_1 & 2 f_3 \\
\end{array}\right] 
+\frac{c}{2}
\left[\begin{array}{ccc}
0 &  f_3 & - f_2 \\
- f_3 &0 &  f_1 \\
 f_2 &- f_1 &0 \\
\end{array}\right] \nn\\
&+\frac{b^{'}}{3}
\left[\begin{array}{ccc}
2 f^{'}_1 &- f^{'}_3& - f^{'}_2 \\
- f^{'}_3 & 2 f^{'}_2 & - f^{'}_1 \\
- f^{'}_2 &- f^{'}_1 & 2 f^{'}_3 \\
\end{array}\right] 
+\frac{c^{'}}{2}
\left[\begin{array}{ccc}
0 &  f^{'}_3 & - f^{'}_2 \\
- f^{'}_3 &0 &  f^{'}_1 \\
 f^{'}_2 &- f^{'}_1 &0 \\
\end{array}\right] 
 \end{align}
where 
$Y^{(4)}_{3}\equiv[y'_1,y'_2,y'_3]^T$ in Appendix.


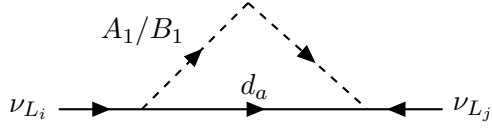
\begin{figure}[tb]
    \begin{tikzpicture}[/tikzfeynman]
    \begin{feynman}
    \vertex (i){$\nu_{L_i}$};
    \vertex[right = 1.5 cm of i](v1);
    \vertex[right = 3. cm of v1](v2);
    \vertex[above right = 2. cm of v1](l1);
    \vertex[right = 1. cm of v2](j){$\nu_{L_j}$};
    \diagram*[large]{
    (i)--[fermion](v1), 
    (v1)--[charged scalar,edge label=$A_1 / B_1$](l1)--[charged scalar](v2), 
    (v1)--[fermion, edge label=$d_a$](v2),
    (v2)--[anti fermion](j)
    };
    \end{feynman}   
    \end{tikzpicture}
  \caption{ One-loop diagrams for generating the neutrino mass matrix.}
  \label{fig:neutrino}
\end{figure}
The charged-lepton sector after spontaneous symmetry breaking is given by
\begin{align}
 - \mathcal{L}_{Y}^\ell = v\frac{h_{ij}}{\sqrt2} \ell_{L_i} e_{R_j}+{\rm h.c.} .
\label{eq:neut} 
\end{align}
Then $m_\ell(\equiv v h/ \sqrt2)$ is diagonalized by two unitary matrices as $D_\ell=V_{\ell_R}^\dag m_\ell V_{\ell_L}$, where $D_\ell
\equiv {\rm diag}(m_e,m_\mu,m_\tau)$ is mass eigenvalues.
Therefore, we find $|D_\ell|^2=V^\dag_{\ell_L} m_\ell^\dag m_\ell V_{\ell_L}$.
The active neutrino mass matrix $m_\nu$
is given at one-loop level through the following interactions:
\begin{align}
 - {L}_{Y}^\ell = 
  F_{aj} \overline{ d'_{R_a}} \nu_{L_j} (c_{a_1} A_1 +s_{a_1} B_1)
-G_{aj} \overline{ d'^c_{L_a}} \nu_{L_j} (-s_{a_1} A_1 + c_{a_1} B_1) ,
\label{eq:neut} 
\end{align}
where $F\equiv V^\dag_{d_R}f$ and  $G\equiv V^T_{d_L}g$ and $d'$ is mass eigenstate.
Then, the neutrino masss matrix in Fig.~\ref{fig:neutrino} is given at one-loop level as follows:
\begin{align}
&(m_{\nu})_{ij}
=
 s_{2a_1}\frac{3}{4(4\pi)^2}
\left[1-\frac{m^2_{A_1}}{m^2_{B_1}}\right]
\sum_{i=1}^3
\left[F^T_{ia} D_{d_a} G_{aj}+ G^T_{ia} D_{d_a} F_{aj} \right] F_I(r_{A_1}, r_{D_{d_i}}),\\
&F_I(r_1, r_2)
=
\frac{r_1(r_2-1)\ln r_1 - r_2(r_1-1)\ln r_2}{(r_1-1)(r_2-1)(r_1-r_2)},\quad (r_1\neq 1),
\end{align}
where  we define $r_{A_1}\equiv (m_{A_1}/m_{B_1})^2$ and  $r_{D_{d_i}}\equiv (D_{d_i}/m_{B_1})^2$.
 $m_\nu$ is diagonalzied by a unitary matrix $V_{\nu}$; $D_\nu\equiv V_{\nu}^T m_\nu V_{\nu}$.
Here, we define a modified neutrino mass matrix as $\tilde m_\nu \equiv m_\nu / s_{2a_1} $.
Then, we rewrite this diagonalization in terms of the modified form $\tilde D_\nu\equiv V_{\nu}^T \tilde m_\nu V_{\nu}$.
Thus, we fix $s_{2a_1}$ by
\begin{align}
({\rm NH}):\  s_{2a_1}^2= \frac{|\Delta m_{\rm atm}^2|}{\tilde D_{\nu_3}^2-\tilde D_{\nu_1}^2},
\quad
({\rm IH}):\  s_{2a_1}^2= \frac{|\Delta m_{\rm atm}^2|}{\tilde D_{\nu_2}^2-\tilde D_{\nu_3}^2},
 \end{align}
where $\tilde m_\nu$ is diagonalized by $V^\dag_\nu (\tilde m_\nu^\dag
\tilde m_\nu)V_\nu=(\tilde D_{\nu_1}^2,\tilde D_{\nu_2}^2,\tilde
D_{\nu_3}^2)$  and $\Delta m_{\rm atm}^2$ is the atmospheric neutrino
mass-squared difference. Here, NH and IH respectively stand for the normal hierarchy
and the inverted hierarchy.  Subsequently, the solar neutrino mass-squared difference is depicted in terms of $s_{2a_1}$ as
follows:
\begin{align}
\Delta m_{\rm sol}^2= {s_{2a_1}^2}({\tilde D_{\nu_2}^2-\tilde D_{\nu_1}^2}).
 \end{align}
This should be within the experimental value, where we adopt NuFit 5.0~\cite{Esteban:2020cvm} to our numerical analysis later. 
 The neutrinoless double beta decay is also given by 
\begin{align}
\langle m_{ee}\rangle=s_{2a_1}|\tilde D_{\nu_1} \cos^2\theta_{12} \cos^2\theta_{13}+\tilde D_{\nu_2} \sin^2\theta_{12} \cos^2\theta_{13}e^{i\alpha_{2}}+\tilde D_{\nu_3} \sin^2\theta_{13}e^{i(\alpha_{3}-2\delta_{CP})}|,
\end{align}
which  may be able to observed by KamLAND-Zen in future~\cite{KamLAND-Zen:2016pfg}. 
The observed mixing matrix of lepton sector~\cite{Maki:1962mu} is given by $V_{\rm PMNS}\equiv V^\dag_{\ell_L} V_{\nu}$.

\section{Numerical analysis \label{sec:numerical}}

Here, we perform numerical analysis.
Before searching for allowed region, we fix some mass parameters as $m_{A_2}=m_{A_1}$ 
and $m_{B_2}=m_{\delta}=m_{B_1}$, {where we require degenerate masses for the components of $\eta$ and $\Delta$ to suppress the oblique parameters $\Delta S$ and $\Delta T$}.
Notice here that our theoretical parameters $a_{u,c,t}, a_{d,s,b}, a_\ell, b_\ell,c_\ell$ are used to determined the experimental masses for quarks and charged-leptons. Thus, only the following input parameters are randomly selected in the range of 
\begin{align}
& (m_{A_1},m_{B_1}) \in [1\,, 100\,]\ \text{TeV},\nn\\
& |g_{u1,u2,u3}| \in \left[10^{-5},\ 1.5\,\right] ,\quad
(|a_\eta|, |b_\eta|, |c_\eta|,  |a|, |b|, |c|, |b'|, |c'|) \in \left[10^{-5},\ 10\,\right].
\label{range_scanning}
\end{align}

Above the range, we have numerical analysis in cases for quark and lepton, where experimental data in the quark sector should be within the range at 3$\sigma$. 
While the one in the lepton sector is discussed in the range within 3$\sigma$(yellow dots) and 5$\sigma$(red dots) applying $\chi^2$ analysis in Nufit 5.0.

\subsection{NH}
\begin{figure}[h]
\begin{minipage}[]{0.4\linewidth}
	\vspace{0mm}
	\includegraphics[{width=\linewidth}]
	{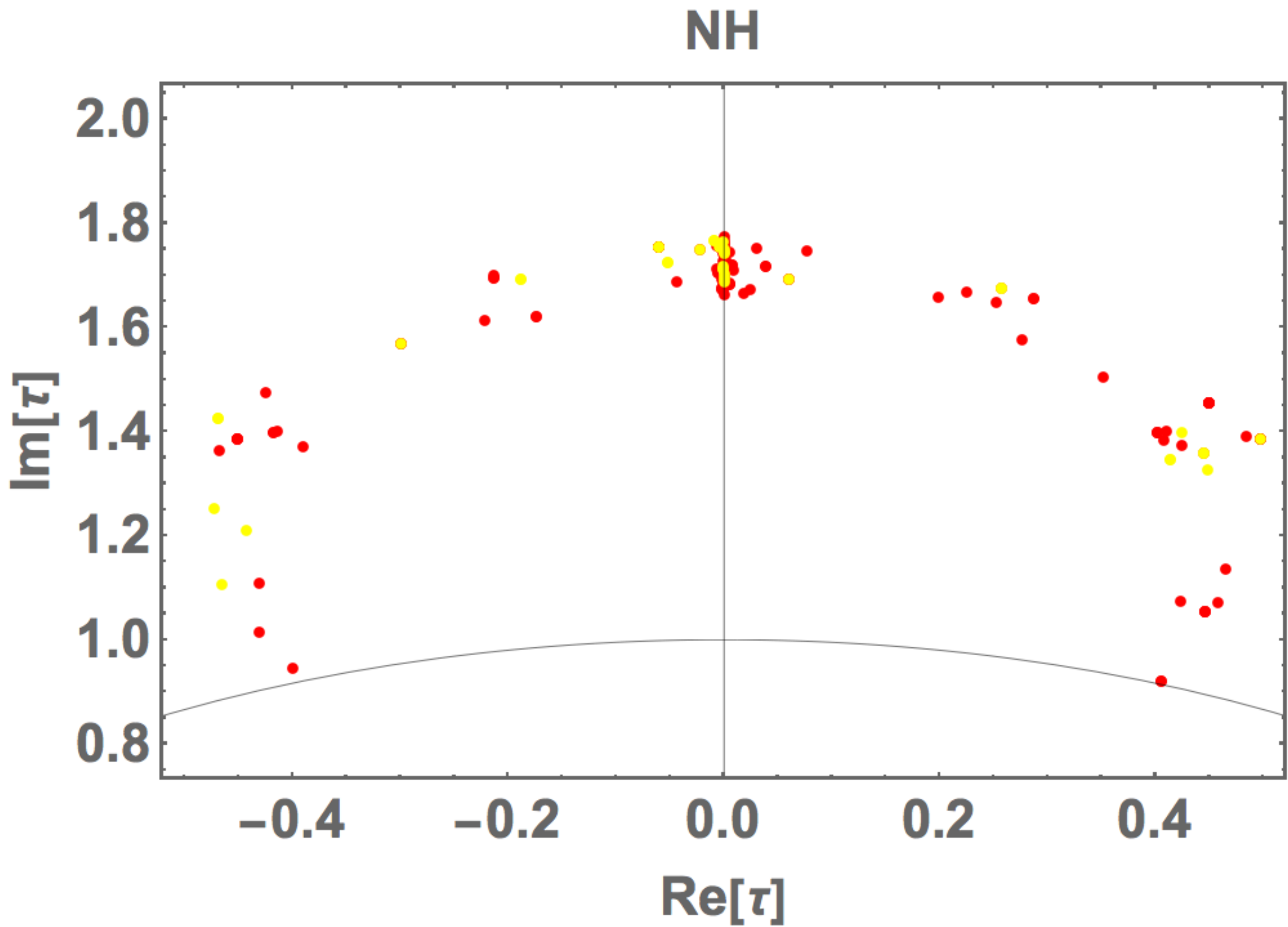}
	\caption{Allowed value of $\tau$. Yellow and red points present the values within 3 and 5$\sigma$.}
	\label{tau_LQMA4_nh}
\end{minipage}
\hspace{5mm}
\begin{minipage}[]{0.4\linewidth}
	\vspace{2mm}
	\includegraphics[{width=\linewidth}]{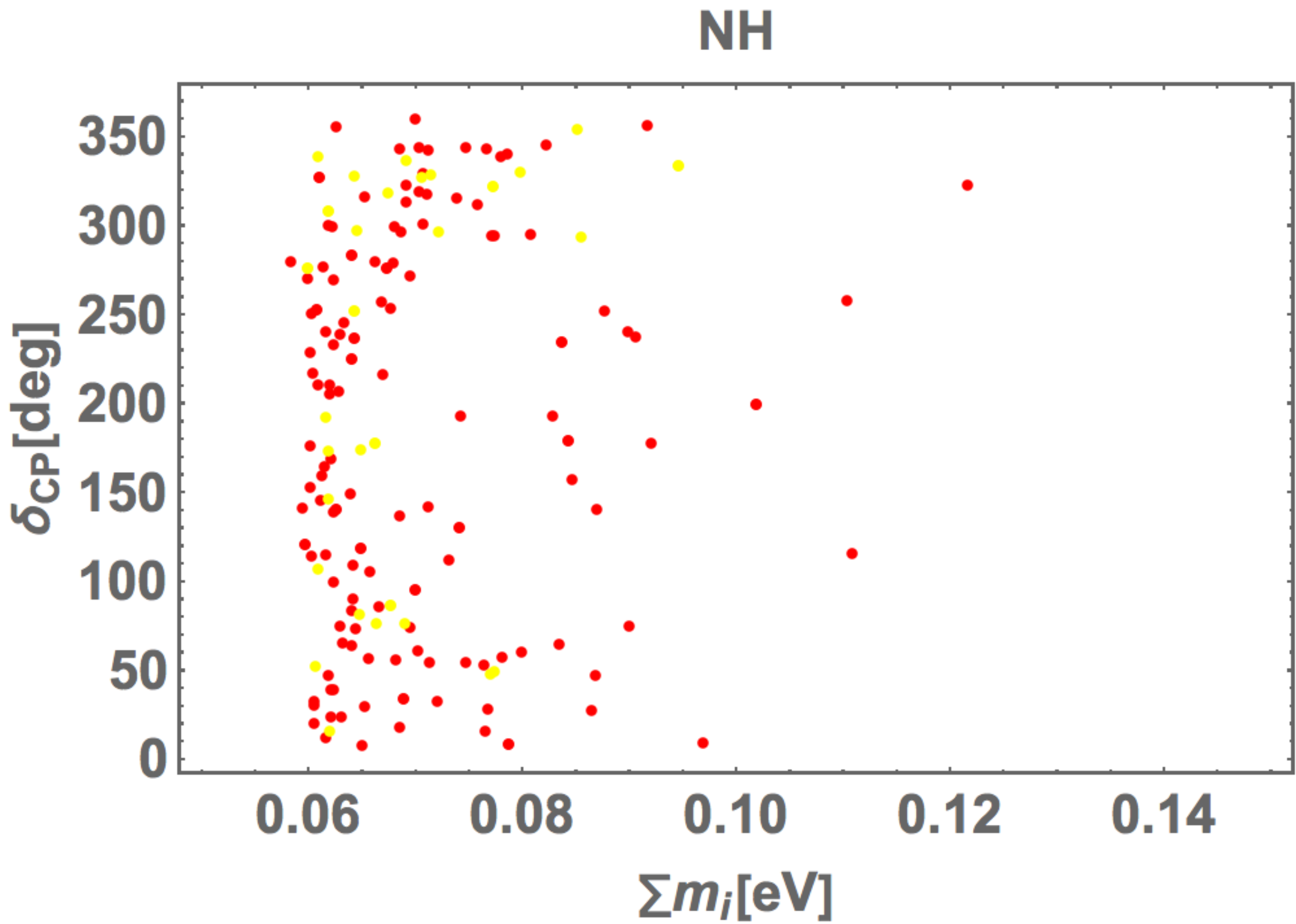}
	\caption{Allowed region of $\delta_{CP}$ in terms of $\sum m_i$.}
	\label{sum-dcp_LQMA4_nh}
\end{minipage}
\hspace{5mm}	
\begin{minipage}[]{0.4\linewidth}
	\vspace{2mm}
	\includegraphics[{width=\linewidth}]{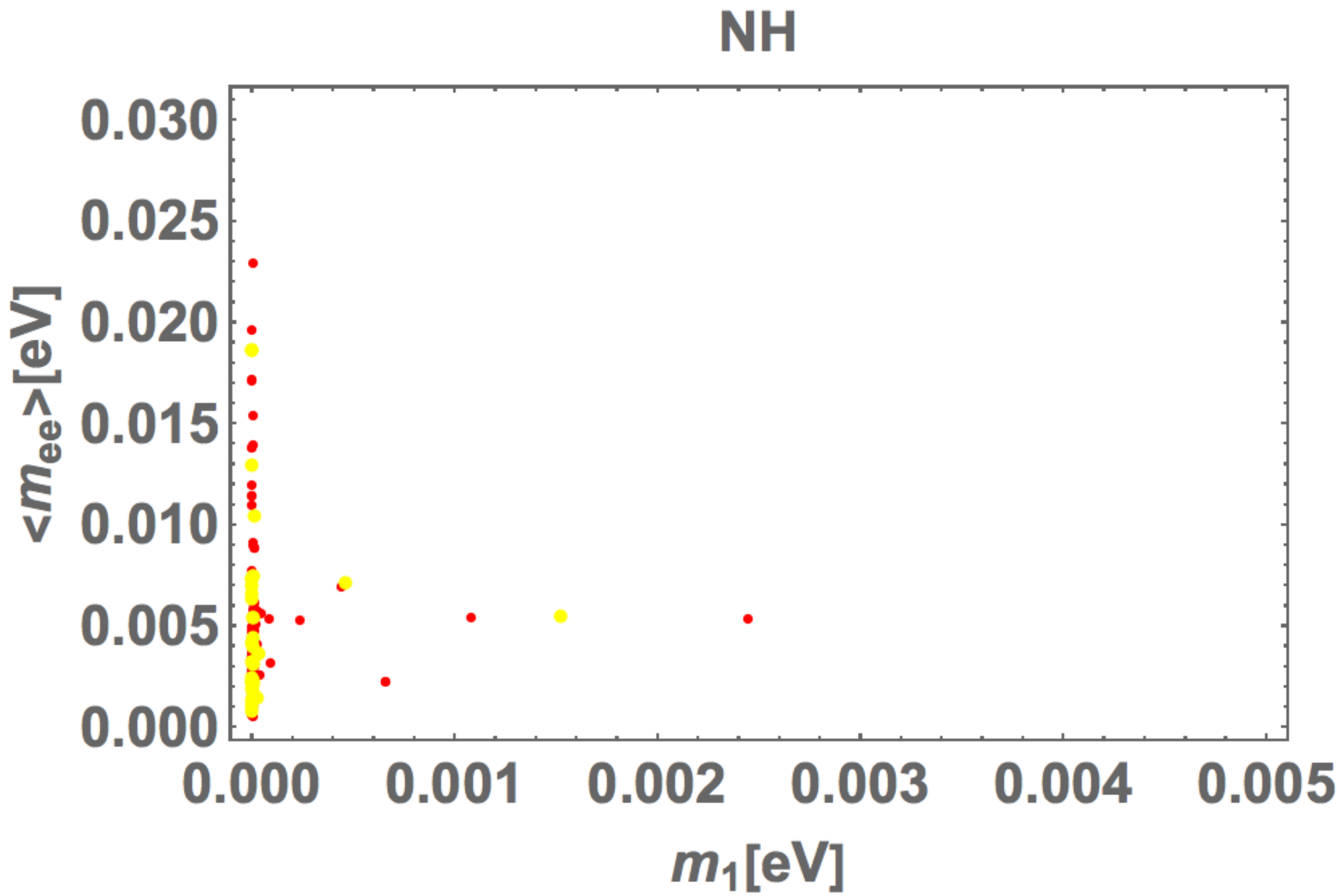}
	\caption{Allowed region of the mass of neutrinoless double beta decay in terms of the lightest neutrino mass.}
	\label{m1-mee_LQMA4_nh}
\end{minipage}
\hspace{5mm}	
\begin{minipage}[]{0.4\linewidth}
	\vspace{2mm}
	\includegraphics[{width=\linewidth}]{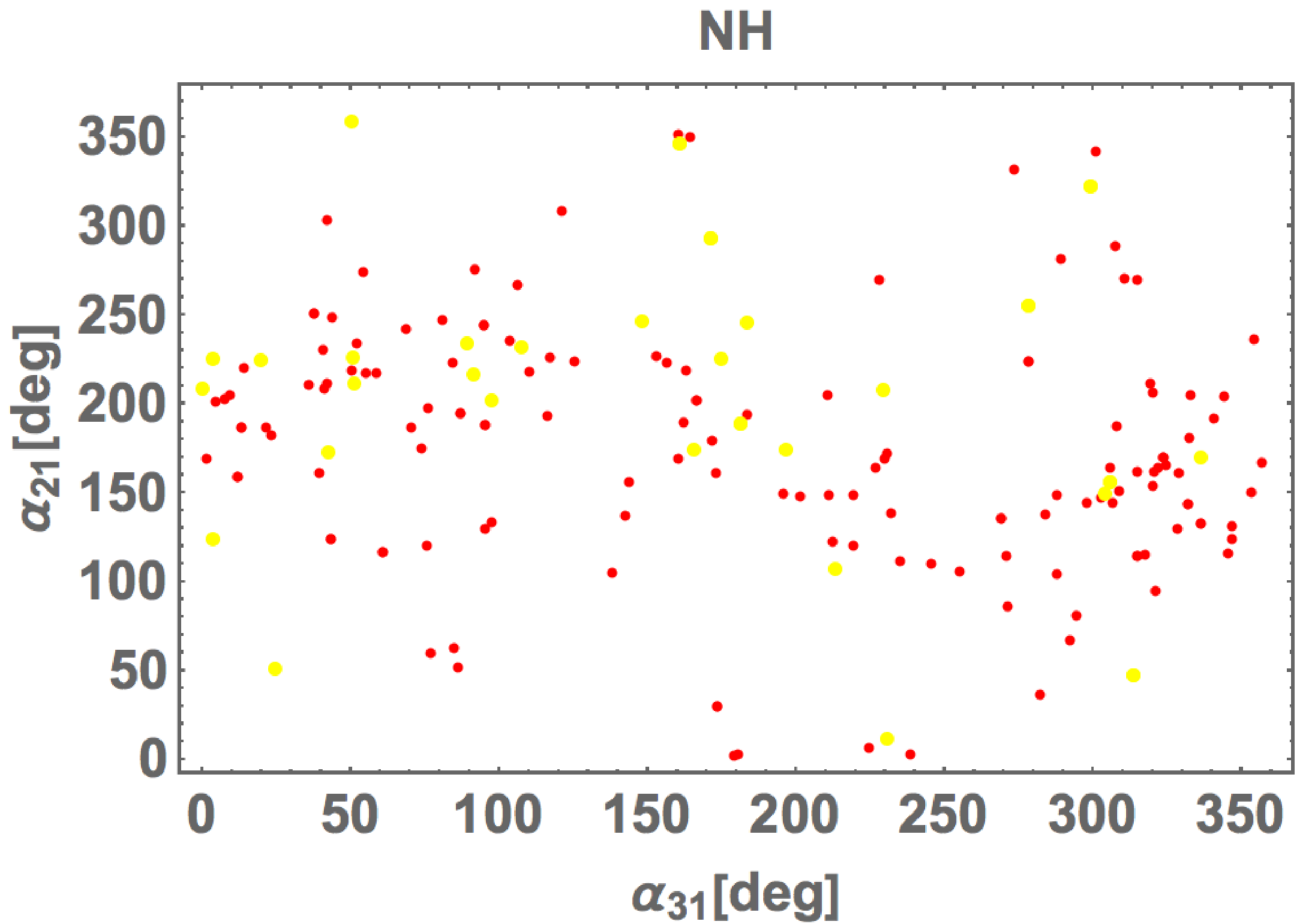}
	\caption{Allowed region of Majorana phases $\alpha_{21}$ in terms of $\alpha_{31}$. }
	\label{majo_LQMA4_nh}
\end{minipage}
\end{figure}

For NH case, we show our results of lepton sector in Figs.~\ref{tau_LQMA4_nh},~\ref{sum-dcp_LQMA4_nh}, \ref{m1-mee_LQMA4_nh}, \ref{majo_LQMA4_nh}.
In Fig.~\ref{tau_LQMA4_nh}, allowed value of $\tau$ is shown where yellow(red) points present the values within 3(5)$\sigma$.
One finds that allowed space is rather localized. Especially, the region at nearby $\tau\sim1.75i$ would be interesting since it is close to the fixed point that has a remnant $Z_3$ symmetry.
In Fig.~\ref{sum-dcp_LQMA4_nh}, we demonstrate allowed region of $\delta_{CP}$ in terms of $\sum m_i$.
$\sum m_i$ is rather localized at $0.06-0.08$ eV while whole the region is allowed for $\delta_{CP}$.
Moreover, almost all the points are within the cosmological constraint $\sim0.12$ eV~\cite{pdg}.
In Fig.~\ref{m1-mee_LQMA4_nh}, we present allowed region of neutrinoless double beta decay $\langle m_{ee}\rangle$ in terms of the lightest neutrino mass $m_1$.
$\langle m_{ee}\rangle$ is allowed up to $0.025$ eV while $m_1$ is allowed up to $0.0025$ eV.
Moreover, allowed region of $m_1$ is localized at around $10^{-6}$ eV indicating tiny mass of the lightest neutrino mass.
In Fig.~\ref{majo_LQMA4_nh}, we depict allowed region of Majorana phases $\alpha_{21}$ and $\alpha_{31}$. 
Both are allowed for whole the region but there would be tendency that $\alpha_{21}$ is localized at around 180$^\circ$.

\begin{figure}[t]
\begin{minipage}[]{0.4\linewidth}
	\vspace{0mm}
	\includegraphics[{width=\linewidth}]
	{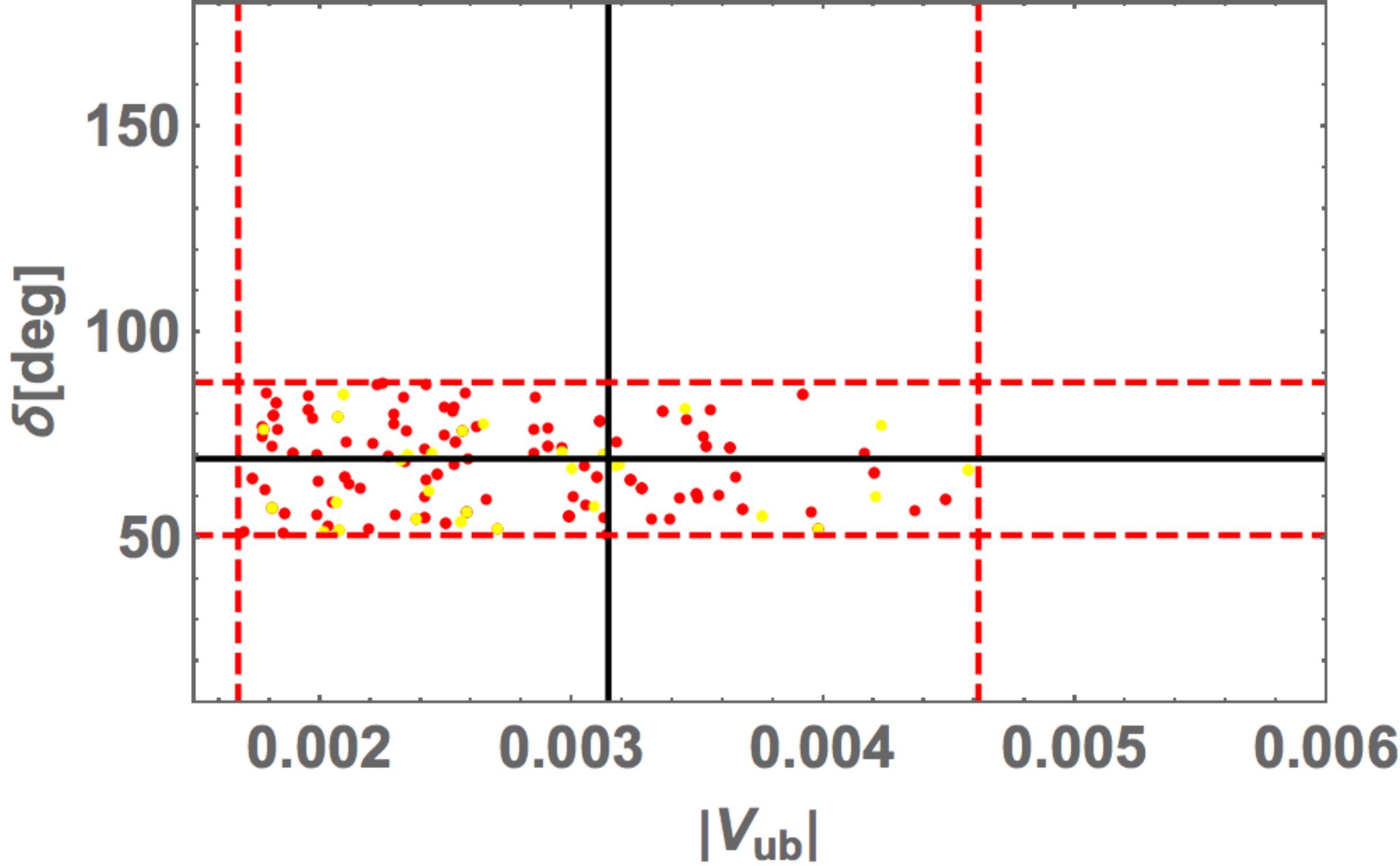}
	\caption{The CP phase of quark $\delta$ versus (1, 3) component of CKM matrix.
	The red dashed lines represent 3$\sigma$ experimental bounds. }
	\label{vub-dcp-nh}
\end{minipage}
\hspace{5mm}
\begin{minipage}[]{0.4\linewidth}
	\vspace{2mm}
	\includegraphics[{width=\linewidth}]{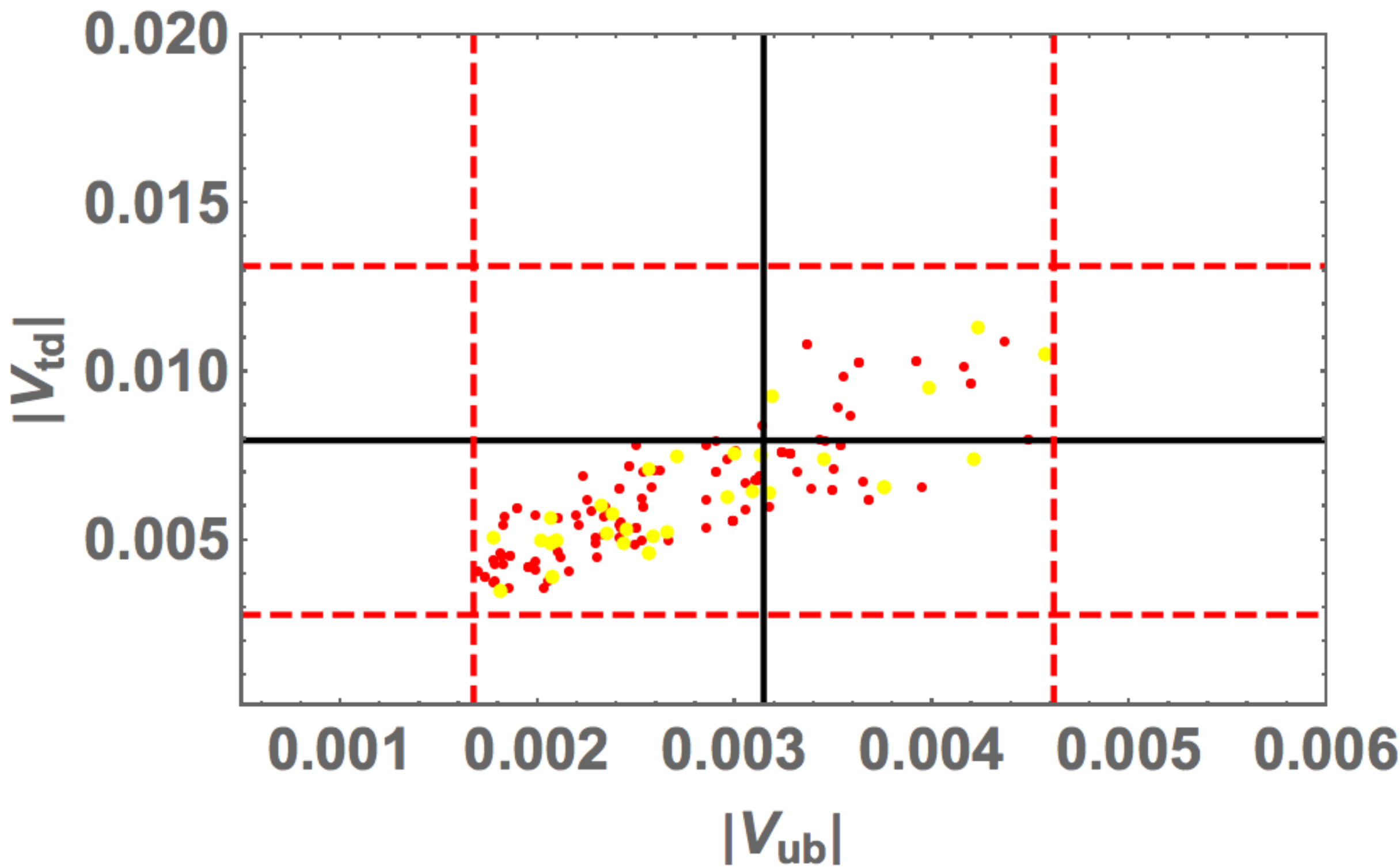}
	\caption{$|V_{ub}|$ versus $|V_{td}|$.}
	\label{vub-vtd-nh}
\end{minipage}
\hspace{5mm}	
\begin{minipage}[]{0.4\linewidth}
	\vspace{2mm}
	\includegraphics[{width=\linewidth}]{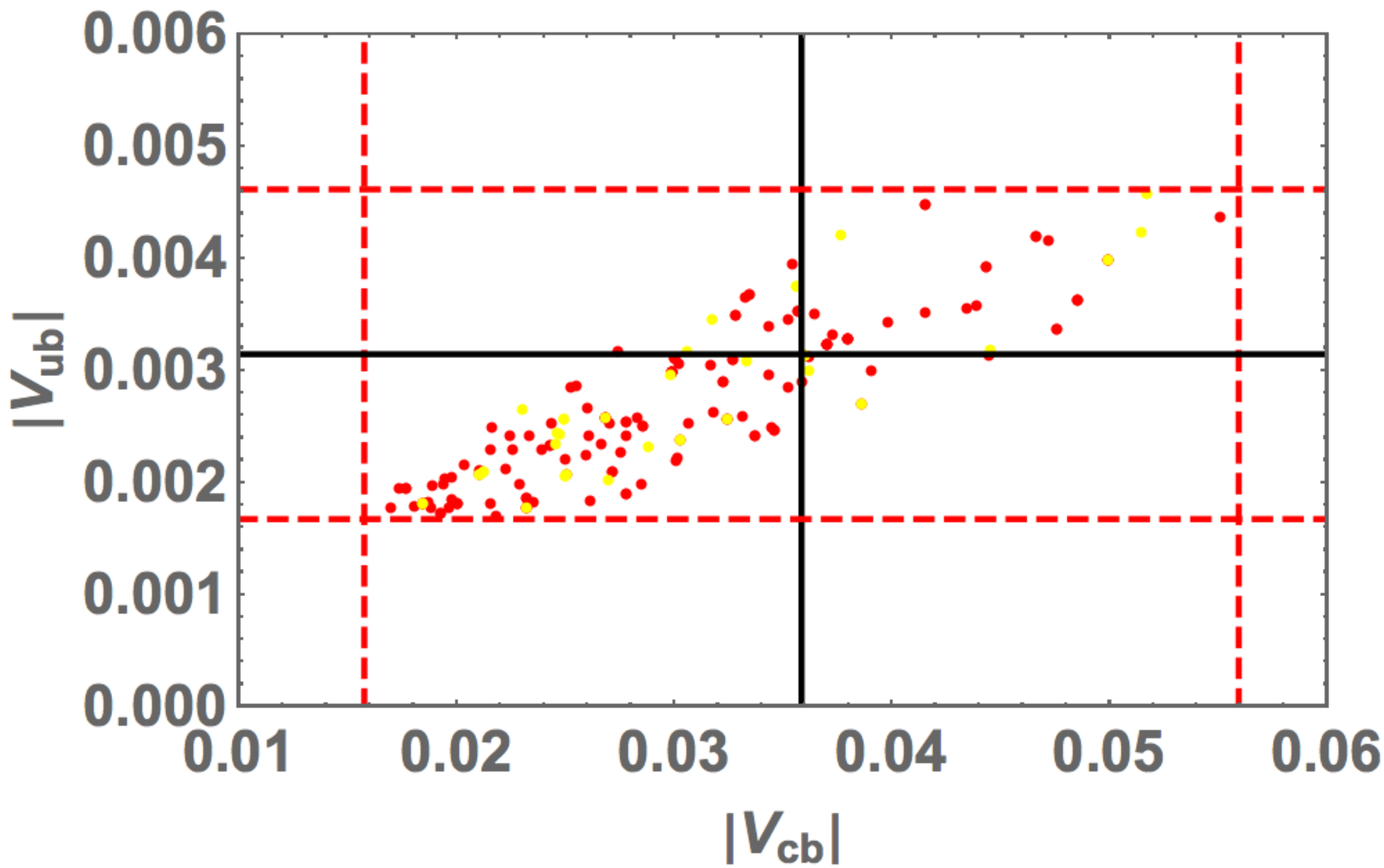}
	\caption{$|V_{cb}|$ versus $|V_{ub}|$.}
	\label{vcb-vub-nh}
\end{minipage}
\end{figure}

In addition to the lepton sector,
we will search for our allowed region of quark sector in Figs.~\ref{vub-dcp-nh}, \ref{vub-vtd-nh}, \ref{vcb-vub-nh}.
Here, the dotted red line at 3$\sigma$ interval while the black line is best fit value.
And the yellow(red) points correspond to 3(5)$\sigma$ of the lepton sector where $\tau$ is commonly used.

In Fig.~\ref{vub-dcp-nh}, we show the CP phase of quark $\delta$ in term of (1, 3) component of CKM matrix; $|V_{ub}|$,
and find whole the region is allowed  at 3$\sigma$ interval.
In Fig.~\ref{vub-vtd-nh}, we show $|V_{ub}|$ and $|V_{td}|$,
and found that there is a weak linearly correlation between them.  
In Fig.~\ref{vcb-vub-nh}, we show $|V_{cb}|$ and $|V_{ub}|$,
and find that there is also a weak linear correlation between them.

{\it Bench mark point for NH}:
We also give a benchmark point to satisfy the quark and lepton masses and mixings as well as phases in the left sides of Tables~\ref{nh-bp-L} and~\ref{nh-bp-Q}, where we extracted a value at nearby $\tau=1.75i$.
The corresponding lepton and neutrino mixings are given by
\begin{align}
V_{\ell_L}&=
\left[\begin{array}{ccc}
-0.75 + 0.00017 i & 0.35 - 0.000026 i & -0.56 \\
0.20 - 0.000012 i & -0.68 + 0.000074 i & -0.70 -  0.000040 i \\
-0.63 + 0.00013 i & -0.64 + 0.000077 i & 0.44 +  0.000044 i \\
\end{array}\right] , \\
V_{\nu}&=
\left[\begin{array}{ccc}
-0.91 - 0.20 i & 0.26 - 0.075 i & -0.22 - 0.027 i \\
0.23 + 0.018 i & -0.045 - 0.28 i & -0.87 - 0.33 i \\
-0.13 + 0.24 i &-0.25 + 0.88 i & -0.25 - 0.15 i \\
\end{array}\right]. 
\end{align}
And the quark mixings are given by
\begin{align}
V_{u_L}&=
\left[\begin{array}{ccc}
-0.75 - 0.057 i & 0.41 + 0.24 i & -0.46 + 0.0034 i \\
-0.62 + 0.017 i & -0.31 - 0.068 i & 0.71 -  0.0045 i \\
0.19 - 0.071 i & 0.76 + 0.30 i & 0.53 - 0.0015 i \\
\end{array}\right] , \\
V_{d_L}&=
\left[\begin{array}{ccc}
-0.64 + 0.000077 i & -0.63 + 0.00013 i & -0.44 - 0.000044 i  \\
-0.68 + 0.000074 i & 0.20 - 0.000012 i & 0.70 + 0.000040 i \\
0.35 - 0.000026 i & -0.75 + 0.00017 i & 0.56  \\
\end{array}\right] .
\end{align}

\begin{table}[h]
	\centering
	\begin{tabular}{|c|c||c|c|} \hline 
			\rule[14pt]{0pt}{0pt}
Lepton		&  NH($\tau\approx 1.75 i$) & IH($\tau\approx 1.06 i$)  & IH($\tau\approx 1.76 i$)\\  \hline
			\rule[14pt]{0pt}{0pt}
		$\tau$&   $  -0.0000945 + 1.75   i$ & $-0.000689 + 1.06  i$  & $-0.000829+ 1.76 i$\\ 
		\rule[14pt]{0pt}{0pt}
		$a_\eta$ &$-0.23 - 1.4 i$ & $-0.31 + 0.013 i $ & $-4.1 + 4.3 i $ \\
		\rule[14pt]{0pt}{0pt}
		$b_\eta$  &  $-0.38 + 1.3 i$ & $-0.045 - 0.027 i$ & $-0.0014 + 0.0032 i $\\
		\rule[14pt]{0pt}{0pt}
		$c_\eta$  &  $0.0077 - 0.031 i $ & $0.0014 - 0.000047 i $ & $-0.0023 + 0.0035 i $\\
		\rule[14pt]{0pt}{0pt}
		$a$  &  $0.00016 + 0.00011 i$ & $3.0 + 0.98 i$ & $0.0085 + 0.025 i $\\
		\rule[14pt]{0pt}{0pt}
		$b$  &  $-0.017 + 0.0013 i$ & $0.00014 + 0.00015 i $ & $-0.096 + 0.041 i $\\
		\rule[14pt]{0pt}{0pt}
		$c$  &  $-0.00030 - 0.00010 i $ & $0.0056 - 0.0021 i $ & $-1.8 + 4.5 i $\\
		\rule[14pt]{0pt}{0pt}
		$b'$  &  $0.00014 - 0.000016 i $ & $0.000024 + 0.000016 i $ & $-0.00011 + 0.00011 i $\\
		\rule[14pt]{0pt}{0pt}
		$c'$  &  $-0.21 - 0.27 i $ & $-0.15 + 0.031 i $ & $-0.0000034 - 0.000060 i $\\
		\rule[14pt]{0pt}{0pt}
		$[\alpha_e, \beta_e,\gamma_e]$ & $[0.0002,9.3\times10^{-7},0.003]$ &$[0.0005,10^{-5},0.007]$  & $[9.1\times10^{-7},1.9\times10^{-4},0.003]$\\
		\rule[14pt]{0pt}{0pt} 
		$\sin^2\theta_{12}$ & $ 0.32$& $0.28$ & $0.33$\\
		\rule[14pt]{0pt}{0pt}
		$\sin^2\theta_{23}$ &  $ 0.56$& $0.46$ & $0.58$\\
		\rule[14pt]{0pt}{0pt}
		$\sin^2\theta_{13}$ &  $ 0.024$&$0.024$ & $0.022$\\
		\rule[14pt]{0pt}{0pt}
		$\delta_{CP}^\ell$ &  $328^\circ$& $ 170^\circ$ & $335^\circ$\\
		\rule[14pt]{0pt}{0pt}
		$[\alpha_{21},\,\alpha_{31}]$ &  $[169^\circ,\, 336^\circ]$ & $[ 167^\circ,\, 159^\circ]$ &  $[ 157^\circ,\, 130^\circ]$	\\	
		\rule[14pt]{0pt}{0pt}
		$\sum m_i$ &  $0.071$\,eV &	 $0.11$\,eV & $0.11$\, eV \\
		\rule[14pt]{0pt}{0pt}
		$s_{2a_1}$ &  $4.6\times10^{-9}$ &	 $5.7\times10^{-5}$ & $1.9\times10^{-9}$ \\
		\rule[14pt]{0pt}{0pt}
		$\langle m_{ee} \rangle$ &  $3.2$\,meV& $21$\,meV  & $19$\,meV \\
		\rule[14pt]{0pt}{0pt}
		$[m_{A_1},m_{B_1}]$ &  $[18,6.0]$\,TeV & $[37, 37]$\,TeV  & $[33, 39]$\,TeV \\
		\rule[14pt]{0pt}{0pt}
		$\sqrt{\chi^2}$ &  $2.9$ & $4.8$  & $4.5$\\
		\hline
	\end{tabular}
	\caption{Numerical values of parameters and observables
		at the best fit points of NH and IH.}
	\label{nh-bp-L}
\end{table}

\begin{table}[h]
	\centering
	\begin{tabular}{|c|c||c|c|} \hline 
			\rule[14pt]{0pt}{0pt}
Quark		&  NH($\tau\approx 1.75 i$) & IH($\tau\approx 1.06 i$)  & IH($\tau\approx 1.76 i$)\\  \hline
			\rule[14pt]{0pt}{0pt}
		$\tau$&   $  -0.0000945 + 1.75 i $ & $-0.000689 + 1.06 i $  & $-0.000829+ 1.76 i $\\ 
		\rule[14pt]{0pt}{0pt}
		$a_u$ &$1.3\times 10^{-7}$ & $8.2\times 10^{-8}$ & $1.1\times 10^{-5}$ \\
		\rule[14pt]{0pt}{0pt}
		$a_c$  &  $4.6\times 10^{-5}$ & $4.2\times 10^{-5}$ & $0.0007$\\
		\rule[14pt]{0pt}{0pt}
		$a_t$  &  $0.017 $ & $0.017$ & $0.22$\\
		\rule[14pt]{0pt}{0pt}
         $g_{u1}$ &$0.00066 + 0.0016 i $ & $-0.00013 + 0.013 i$ & $0.57 - 0.35 i $ \\
		\rule[14pt]{0pt}{0pt}
		$g_{u2}$  &  $0.053 + 0.30 i $ & $0.094 + 0.24 i $ & $-0.060 - 0.41 i $\\
		\rule[14pt]{0pt}{0pt}
		$g_{u3}$  &  $0.11 + 0.0061 i $ & $0.091 + 0.018 i $ & $0.046 + 0.021 i $\\
		\rule[14pt]{0pt}{0pt}
		$a_d$  &  $0.00018$ & $0.00014$ & $0.00047$\\
		\rule[14pt]{0pt}{0pt}
		$a_s$  &  $1.2 \times 10^{-5} $ & $8.7\times 10^{-6}$ & $5.2\times 10^{-5}$\\
		\rule[14pt]{0pt}{0pt}
		$a_b$  &  $0.011$ & $0.011$ & $0.025$\\
		\rule[14pt]{0pt}{0pt}
		$|V_{us}|$ & $ 0.23$& $0.22$ & $0.22$\\
		\rule[14pt]{0pt}{0pt}
		$|V_{cb}|$ &  $ 0.033$& $0.027$ & $0.042$\\
		\rule[14pt]{0pt}{0pt}
		$|V_{ub}|$ &  $ 0.0031$&$0.0020$ & $0.0039$\\
		\rule[14pt]{0pt}{0pt}
		$\delta_{CP}$ &  $58^\circ$& $ 51^\circ$ & $83^\circ$\\
		\hline
	\end{tabular}
	\caption{Numerical values of parameters and observables
		at the best fit points of NH and IH.}
	\label{nh-bp-Q}
\end{table}

\subsection{IH}
In case of IH, we obtain less allowed parameter points compared to the case of $NH$ since it is more difficult to fit the data.
Since there are no points within 3$\sigma$ region but few points within  5$\sigma$ region, we will explain the tendency instead of showing scattering plots.
The value of $\tau$ is interestingly localized at nearby two fixed points $i$ and $1.74i$, each of which has remnant symmetry of $Z_2$ and $Z_3$.
%
$\sum m_i$ is localized at $0.10 - 0.12$ eV while  $\delta_{CP}$ is allowed for the range of $150^\circ - 360^\circ$.
Moreover, almost all the points are within the cosmological constraint $\sim0.12$ eV that is similar to the NH case.
%
$\langle m_{ee}\rangle$ is localized at around $0.016 - 0.024$ eV while $m_1$ is allowed up to $1.2\times10^{-4}$ eV.
Moreover,  $m_1$ is also localized at around $10^{-6}$ eV .
%
$\alpha_{21}$ is localized at around 180$^\circ$,while $\alpha_{21}$ is allowed for the range of $100^\circ -360^\circ$.

In addition to the lepton sector,
we discuss our allowed region of quark sector. 
Even though the allowed points are not so many, we might say something from our analysis as follows.
As for the CP phase of quark $\delta$ in term of (1, 3) component of CKM matrix; $|V_{ub}|$, 
we found whole the region is allowed  at 3$\sigma$ interval.
As for $|V_{ub}|$ and $|V_{td}|$,
we find that there is a weak linearly correlation between them. 
As for $|V_{cb}|$ and $|V_{ub}|$,
we find that there is also a weak linear correlation between them.

{\it Bench mark point for IH}:
We give two interesting benchmark points; $\tau\approx 1.06 i,\ 1.76 i$  to satisfy the quark and lepton masses and mixings as well as phases in the center and right sides of Tables~\ref{nh-bp-L} and~\ref{nh-bp-Q}.
The lepton and neutrino mixings are given by
\begin{align}
\tau\approx 1.06 i:&\nn\\
V_{\ell_L}&=
\left[\begin{array}{ccc}
-0.65 + 0.0068 i & 0.72 + 0.00024 i & -0.25 + 0.00061 i \\
-0.47 + 0.0067 i & -0.64 + 0.0012 i &-0.61 + 0.00072 i \\
-0.60 + 0.0068 i & -0.28 + 0.00098 i & 0.75 + 0.000042 i \\
\end{array}\right] , \\
V_{\nu}&=
\left[\begin{array}{ccc}
-0.63 + 0.12 i &0.053 - 0.029 i & -0.13 - 0.75 i \\
0.090 + 0.11 i &0.14 + 0.98 i & 0.015 - 0.089 i \\
-0.75 + 0.10 i &0.12 + 0.097 i & 0.068 + 0.63 i \\
\end{array}\right] , \\
\end{align}
\begin{align}
\tau\approx 1.76 i:&\nn\\
V_{\ell_L}&=
\left[\begin{array}{ccc}
-0.21 - 0.000031 i & 0.80 - 0.0015 i & 0.56 + 0.000052 i \\
0.63 - 0.00065 i & -0.33 + 0.00026 i &0.70 + 0.00032 i \\
0.75 - 0.00091 i &0.50 - 0.0010 i &-0.44 - 0.00036 i \\
\end{array}\right] , \\
V_{\nu}&=
\left[\begin{array}{ccc}
-0.010 + 0.016 i &0.063 - 0.098 i &-0.87 + 0.48 i \\
0.28 - 0.037 i &-0.77 + 0.56 i & -0.12 + 0.014 i \\
0.53 + 0.80 i &0.25 + 0.13 i & 0.016 + 0.0094 i \\
\end{array}\right] .
\end{align}
The quark mixings are given by
\begin{align}
\tau\approx 1.06 i:&\nn\\
V_{u_L}&=
\left[\begin{array}{ccc}
-0.59 + 0.26 i & -0.18 - 0.067 i & -0.71 - 0.23 i \\
-0.55 + 0.21 i & -0.50 - 0.030 i & 0.61 + 0.18 i \\
-0.41 + 0.28 i & 0.84 - 0.062 i & 0.20 + 0.080 i \\
\end{array}\right] , \\
V_{d_L}&=
\left[\begin{array}{ccc}
-0.60 + 0.0068 i & -0.28 + 0.00096 i & -0.75 - 0.000044 i \\
-0.47 + 0.0067 i & -0.64 + 0.0011 i & 0.61 - 0.00072 i \\
-0.65 + 0.0068 i & 0.72 + 0.00022 i & 0.25 - 0.00061 i \\
\end{array}\right] , 
\end{align}
\begin{align}
\tau\approx 1.76 i:&\nn\\
V_{u_L}&=
\left[\begin{array}{ccc}
-0.76 - 0.042 i & 0.43 + 0.19 9 & -0.45 + 0.018 8i\\
-0.62 + 0.015 i & -0.33- 0.052 i & 0.71 - 0.026 i \\
0.19 - 0.054 i & 0.78 + 0.23 i & 0.54 - 0.014 i \\
\end{array}\right] , \\
V_{d_L}&=
\left[\begin{array}{ccc}
-0.64 + 0.00066 i & -0.63 + 0.0011 i & -0.44 - 0.00039 i \\
-0.68 + 0.00064 i & 0.21 - 0.00013 i & 0.70 + 0.00037 i \\
0.35 - 0.00023 i & -0.75 + 0.0015 i & 0.56 + 0.000089 i \\
\end{array}\right].
\end{align}

\section{Conclusions}
We have proposed a LQ model to explain the masses and mixings for quark and lepton,
introducing modular $A_4$ symmetry.
Due to nature of LQ model that lepton(quark) directly connects to the quark(lepton) via LQ, a single modulus number 
has to be applied that leads to a good motivation towards unification of quark and lepton flavor in $A_4$ modular symmetry. 
After giving an assignment for quark sector to reproduce the experimental results at 3$\sigma$ interval, we have also constructed the lepton sector, where
the neutrino mass matrix is induced at one-loop level running down quark sector, unified value of $\tau$ is used for quark and lepton.

Then, we have performed numerical analysis to search for allowed region satisfying experimental measurements for both quark and lepton sector, depending on NH and IH.
In case of NH, we have found  rather wide allowed space within 3$\sigma$ interval and obtained tendency of observables for quark and lepton.
Especially, we have found allowed region at nearby $\tau=1.75 i$ that is close to the fixed point of $\tau=i\infty$.
Thus, we have also shown a promising bench mark point at around the solution.

In case of IH, we would not found the allowed region within 3$\sigma$ interval, but found within 5$\sigma$ interval.
Although the number of allowed point is few, we have found all the allowed regions are localized at nearby $\tau=i,\ 1.76i$, both of which are nearby fixed points. We have shown them as benchmark points. These would be tested near future.

\section*{Acknowledgments}
\vspace{0.5cm}
{\it
This research was supported by an appointment to the JRG Program at the APCTP through the Science and Technology Promotion Fund and Lottery Fund of the Korean Government. This was also supported by the Korean Local Governments - Gyeongsangbuk-do Province and Pohang City (H.O.),  European Regional Development Fund-Project Engineering Applications 
of Microworld Physics (Grant No. CZ.02.1.01/0.0/0.0/16\_019/0000766) (Y.O.). H. O. is sincerely grateful for the KIAS member.}

\section*{Appendix}
The  modular forms of weight 2, $Y^{(2)}_{\bf3} = [y_{1},y_{2},y_{3}]^T$,  transforming
as a triplet of $A_4$ is written in terms of Dedekind eta-function  $\eta(\tau)$ and its derivative:
\begin{eqnarray} 
\label{eq:Y-A4}
y_{1}(\tau) &=& \frac{i}{2\pi}\left( \frac{\eta'(\tau/3)}{\eta(\tau/3)}  +\frac{\eta'((\tau +1)/3)}{\eta((\tau+1)/3)}  
+\frac{\eta'((\tau +2)/3)}{\eta((\tau+2)/3)} - \frac{27\eta'(3\tau)}{\eta(3\tau)}  \right), \nonumber \\
y_{2}(\tau) &=& \frac{-i}{\pi}\left( \frac{\eta'(\tau/3)}{\eta(\tau/3)}  +\omega^2\frac{\eta'((\tau +1)/3)}{\eta((\tau+1)/3)}  
+\omega \frac{\eta'((\tau +2)/3)}{\eta((\tau+2)/3)}  \right) , \label{eq:Yi} \\ 
y_{3}(\tau) &=& \frac{-i}{\pi}\left( \frac{\eta'(\tau/3)}{\eta(\tau/3)}  +\omega\frac{\eta'((\tau +1)/3)}{\eta((\tau+1)/3)}  
+\omega^2 \frac{\eta'((\tau +2)/3)}{\eta((\tau+2)/3)}  \right), 
\nonumber\\
\eta(\tau) &=& q^{1/24}\Pi_{n=1}^\infty (1-q^n), \quad q=e^{2\pi i \tau}, \quad \omega=e^{2\pi i /3}
\nonumber
\end{eqnarray}
%
Then, any multiplets of higher weight are constructed by multiplication rules of $A_4$,
and one finds the following :
\begin{align}
&Y^{(4)}_{\bf1}=y^2_1+2y_2y_3,\quad
Y^{(4)}_{\bf3}\equiv
\left[\begin{array}{c}
y'_1 \\ 
y'_2 \\ 
y'_3 \\ 
\end{array}\right]
=
\left[\begin{array}{c}
y^2_1-y_2y_3 \\ 
y^2_3-y_1y_2 \\ 
y^2_2-y_1y_3 \\ 
\end{array}\right], \quad 
Y^{(6)}_{\bf 1}= y_1^2 + y_2^2 + y_3^2 - 3 y_1 y_2 y_3, 
\end{align}
\begin{align}
&Y^{(6)}_{\bf3}\equiv
\left[\begin{array}{c}
f_1 \\ 
f_2 \\ 
f_3 \\ 
\end{array}\right]
=
\left[\begin{array}{c}
y^3_1 + 2 y_1 y_2 y_3 \\ 
y^2_1 y_2 +2 y_2^2 y_3 \\ 
y^2_1 y_3 + 2 y_3^2 y_2 \\ 
\end{array}\right],\quad
Y^{(6)}_{\bf 3'}\equiv
\left[\begin{array}{c}
f'_1 \\ 
f'_2 \\ 
f'_3 \\ 
\end{array}\right]
=
\left[\begin{array}{c}
y^3_3 + 2 y_1 y_2 y_3 \\ 
y^2_3 y_1 +2 y_1^2 y_2 \\ 
y^2_3 y_2 + 2 y_2^2 y_1 \\ 
\end{array}\right].
\end{align}



\end{document}